\documentstyle[preprint,prc,aps]{revtex}
\begin{document}
\title{THE NUCLEON-NUCLEON INTERACTION IN A CHIRAL CONSTITUENT QUARK MODEL}
\author{Fl. Stancu and S. Pepin}
\address{Universit\'{e} de Li\`ege, Institut de Physique B.5, Sart Tilman,
B-4000 Li\`ege 1, Belgium}
\author{L. Ya. Glozman}
\address{Institute for Theoretical Physics, University of Graz, 8010 Graz,
Austria}
\date{\today}
\maketitle
\begin{abstract}
  We study the short-range nucleon-nucleon interaction in a chiral
constituent quark model by diagonalizing a Hamiltonian
comprising a linear confinement and a Goldstone boson exchange interaction
between quarks. The six-quark harmonic oscillator basis contains up to two
excitation quanta.  We show that the highly dominant configuration is $\mid
s^4p^2[42]_O [51]_{FS}>$ due to its specific flavour-spin symmetry.
Using the Born-Oppenheimer approximation we find a strong effective
repulsion at zero separation between nucleons in both $^3S_1$ and $^1S_0$
channels. The symmetry structure of the highly dominant configuration
implies the existence of a node in the S-wave relative motion wave function
at short distances. The amplitude of the oscillation of the wave function
at short range will be however strongly suppressed.
We discuss the mechanism leading
to the effective short-range repulsion within the chiral  constituent quark
model as compared to that related with the one-gluon exchange interaction.
\end{abstract}
\section{Introduction}
 An interest in the constituent quark model (CQM) has recently been
revitalized \cite{GR96} after recognizing the fact that the constituent
(dynamical) mass of the light quarks appears as a direct consequence of the
spontaneous chiral symmetry breaking (SCSB) \cite{W79,MG84}
and is related with the
light quark condensates $<\bar{q}q>$ of the QCD vacuum. This
feature becomes explicit in any microscopical approach to SCSB in QCD, e.g. in
the instanton gas (liquid) model \cite{CDG78}. The mechanism of the dynamical
mass generation in the Nambu-Goldstone mode of chiral symmetry is very
transparent  within the
$\sigma$-model \cite{GL60} or Nambu and Jona-Lasinio model \cite{NJL61}.
Another consequence of the chiral symmetry in the Nambu-Goldstone mode is the
appearance of an octet of Goldstone bosons ($\pi, K, \eta$ mesons). It was
suggested in \cite{GR96} that beyond the scale of SCSB, nonstrange and strange
baryons should be viewed as systems of three constituent quarks which interact
via the exchange of Goldstone bosons and are subject to confinement. This type
of interaction between the constituent quarks provides a very satisfactory
description of the low-lying nonstrange and strange baryon spectra
\cite{GR96,GPP96,GPPVW97} including the correct ordering of the levels with
positive and negative parity in all parts of the considered spectrum.
 
So far, all studies of the short-range NN interaction within the constituent
quark model were based on the one-gluon exchange interaction (OGE) between
quarks. They explained the short-range repulsion in the NN system as due to
the colour-magnetic part of OGE combined with quark interchanges between 3q
clusters. (For reviews and earlier references see \cite{OY84,MW88,S89}). There
are also models which attribute the short-range repulsion in the NN system to
the colour-electric part of OGE \cite{PSKW96}.
 
In order to provide the necessary long- and intermediate-range attraction in
the baryon-baryon system, hybrid models were suggested
\cite{KNO91,ZFSG94,FNS96}, where in addition to OGE, the
quarks  belonging to different 3q clusters interact via pseudoscalar and
scalar meson exchange. In these hybrid models the short-range repulsion in
the NN system is still attributed to OGE between the constituent quarks.
 
It has been shown, however \cite{GR96,G97}, that the hyperfine splittings
as well as the correct ordering of
positive and negative parity states in spectra of baryons with u,d,s quarks are
produced in fact not by the colour-magnetic part of OGE, but by the
short-range part     of the Goldstone boson exchange (GBE) interaction. This
short-range part of GBE has just opposite sign as compared to the Yukawa tail
of the GBE interaction and is much stronger at short interquark separations.
There is
practically no  room for OGE in light baryon spectroscopy and any
appreciable amount of colour-magnetic interaction, in addition to GBE,
destroys the spectrum \cite{GPPVW97}. If so, the question arises which
interquark interaction is responsible for the short-range NN repulsion. The
goal of this paper is to show that the same short-range part of GBE, which
produces good baryon spectra, also induces a short-range repulsion in the NN
system.
 
The present study is rather exploratory. We calculate an effective
NN interaction
at zero separation distance only. We also want to stress that all main
ingredients of the NN interaction, such as the long- and middle-range
attraction and the short-range repulsion are implied by the chiral
constituent quark model. Indeed, the long- and middle-range attraction
automatically appear in the present framework due to the long-range
Yukawa tail of the pion-exchange interaction between quarks belonging to
different nucleons and due to 2$\pi$ (or sigma) exchanges.
Thus, the only important open question is whether or not the chiral
constituent quark model is able to produce a short-range repulsion in
the NN system.
 
For this purpose, we diagonalize the Hamiltonian of Ref.\cite{GPP96} in a
six-quark harmonic oscillator basis up to two excitations quanta. Using
the Born-Oppenheimer (adiabatic) approximation, we obtain an effective
internucleon potential at zero separation between nucleons from the
difference between the lowest eigenvalue and two times the nucleon mass
calculated in the same model. We find a strong effective repulsion between
nucleons in both $^3S_1$ and $^1S_0$ channels of a height of 800-1300
MeV. This repulsion implies a strong suppression of the NN wave function
in the nucleon overlap region as compared to the wave function of the well
separated nucleons.

Due to the specific flavour-spin symmetry of the GBE
interaction, we also find that the highly dominant 6q configuration 
at zero-separation
between nucleons is $|s^4p^2 [42]_O [51]_{FS}>$. As a consequence the
6q region (i.e. the nucleon overlap region) cannot be adequately
represented by the one-channel resonating group method (RGM) ansatz
$\hat{A}\{N(1,2,3)N(4,5,6)\chi(\vec{r})\}$ which is commonly used at present
for the short-range NN interaction with the OGE interaction.
 
The symmetry structure $[42]_O [51]_{FS}$ of the lowest configuration
will induce an additional effective repulsion at short range related
to the ``Pauli forbidden state" in this case. This latter effective
repulsion is not related to the energy of the lowest configuration
as compared to two-nucleon threshold and thus cannot be obtained within
the Born-Oppenheimer approximation procedure. We notice, however, that
the structure of the six-quark wave function in the nucleon overlap
region is very different from the one associated with the soft or hard
core $NN$ potentials.

This paper is organized as follows. In section 2, in a qualitative
analysis at the Casimir operator level, we show that the short-range GBE
interaction generates a repulsion between nucleons in both $^3S_1$ and
$^1S_0$ channels. We also suggest there that the configuration with the
$[51]_{FS}$ flavour-spin symmetry should be the dominant one.
Section 3 describes the Hamiltonian. Section 4 contains results of
the diagonalization of the 6q Hamiltonian and of the NN
effective interaction at zero separation between nucleons. The structure
of the short-range wave function is also discussed in this section. In section
5, we show why the single-channel RGM ansatz is not adequate in the
present case. In
section 6, we present a summary of our study.

\section{A qualitative analysis at the Casimir operator level}
In order to have a preliminary qualitative insight it is convenient
first to consider a schematic model which neglects the radial dependence
of the GBE interaction. In this model the short-range part of the GBE
interaction between the
constituent quarks is approximated by \cite{GR96}
\begin{equation}
 V_{\chi} = - C_{\chi} \sum_{i<j}  \lambda_{i}^{F} . \lambda_{j}^{F}
\vec{\sigma}_i . \vec{\sigma}_j ,
\label{opFS}
\end{equation}
where $\lambda^{F}$ with an implied summation over F (F=1,2,...,8)
and $\vec{\sigma}$ are the quark flavour
Gell-Mann and spin matrices respectively.
The minus sign of the interaction (\ref{opFS}) is
related to the sign of the short-range part of the pseudoscalar
meson-exchange interaction (which is opposite to that of the Yukawa tail),
crucial for the hyperfine splittings in baryon spectroscopy. The
constant $C_{\chi}$ can be determined from the $\Delta-N$ splitting.
For that purpose one only needs the spin (S), flavour (F) and
flavour-spin (FS) symmetries of the $N$ and $\Delta$ states,
identified by the corresponding partitions [f] associated with the
groups $SU(2)_S, SU(3)_F$ and $SU(6)_{FS}$ :
\begin{eqnarray}
|N> &=& |s^3 [3]_{FS} [21]_F [21]_S > , \\
 |\Delta> &=& |s^3 [3]_{FS} [3]_F [3]_S > .
\end{eqnarray}
 Then the matrix elements of the interaction (\ref{opFS}) are \cite{GR96} :
\begin{eqnarray}
<N | V_{\chi} | N> &=& -14 C_{\chi} , \\
<\Delta | V_{\chi} | \Delta> &=& -4 C_{\chi} .
\end{eqnarray}
Hence $E_{\Delta}-E_N = 10 C_{\chi}$, which gives $C_{\chi} = 29.3$MeV, if
one uses the experimental value of 293 MeV for the $\Delta - N$ splitting.
 
To see the effect of the interaction (\ref{opFS}) in the six-quark system,
the most convenient is to use the coupling scheme called FS, where the
spatial $[f]_O$ and colour $[f]_C$ parts are coupled together to $[f]_{OC}$,
and then to the $SU(6)_{FS}$ flavour-spin part of the wave function in order
to provide a totally antisymmetric wave function in the OCFS space
\cite{H81}. The antisymmetry condition requires $[f]_{FS} =
[\tilde{f}]_{OC}$, where
$[\tilde{f}]$ is the conjugate of $[f]$.
 
The colour-singlet 6q state is $[222]_C$. Assuming that N has a $[3]_O$
spatial symmetry, there are two possible states $[6]_O$ and
$[42]_O$ compatible with the S-wave relative motion in the NN system
 \cite{NST77}.
The flavour and spin symmetries are $[42]_F$ and $[33]_S$ for $^1S_0$ and
$[33]_F$ and $[42]_S$ for $^3S_1$ channels. Applying the inner product rules
of the symmetric group for both the $[f]_O \times [f]_C$ and $[f]_F
\times [f]_S$ products one arrives at the following 6q antisymmetric states
associated with the $^3S_1$ and $^1S_0$ channels \cite{H81,book} :
$| [6]_O [33]_{FS} >$, $| [42]_O [33]_{FS} >$, $| [42]_O [51]_{FS} >$,
$| [42]_O [411]_{FS} >$, $| [42]_O [321]_{FS} >$, $| [42]_O [2211]_{FS} >$.
 
Then the expectation values of the GBE interaction (\ref{opFS}) for these
states  can be easily calculated in terms of the Casimir operators
eigenvalues for the groups $SU(6)_{FS}$, $SU(3)_F$ and $SU(2)_S$ using the
formula given in Appendix A. The corresponding matrix elements are given in
Table 1, from where one can see that, energetically, the most favourable
configuration is $[51]_{FS}$. This is a direct consequence of the general
rule that at short range and with fixed spin and flavour,
the more ``symmetric" a given FS Young diagram is, the more negative is
the expectation value of (\ref{opFS}) \cite{GR96}.
The difference in the potential
energy between the configuration $[51]_{FS}$ and $[33]_{FS}$ or $[411]_{FS}$
is of the order :
\begin{equation}
\begin{array}{cccc}
<[33]_{FS} | V_{\chi} | [33]_{FS}> & - & <[51]_{FS} | V_{\chi} |
[51]_{FS}> & = \\
<[411]_{FS} | V_{\chi} | [411]_{FS}> & - & <[51]_{FS} | V_{\chi} |
[51]_{FS}> & = 24 C_{\chi}\\
\end{array}
\end{equation}
and using $C_{\chi}$ given above one obtains approximately 703 MeV
for both the SI = 10 and 01 sectors.
 
In a harmonic oscillator basis containing up to $2\hbar\omega$ excitation
quanta,
there are two different 6q states corresponding to the $[6]_O$ spatial symmetry
with removed center of mass motion. One of them, $|s^6 [6]_O>$, belongs to
the $N=0$ shell, where $N$ is the number of excitation quanta
in the system,  and the other,
$ \sqrt{\frac {5}{6}} |s^52s [6]_O> - \sqrt{\frac {1}{6}} |s^4p^2 [6]_O>$,
belongs to the $N=2$ shell. There is only one state with $[42]_O$ symmetry,
the $|s^4p^2 [42]_O>$ state belonging to the N=2 shell. While here and below
we use notations of
the shell model it is always assumed that the center of mass motion is
removed.
 
The kinetic energy KE for the $|s^4p^2 [42]_O>$ state is larger than the
one for the $|s^6[6]_O>$ state
by $KE_{N=2} - KE_{N=0} = \hbar \omega$. Taking $\hbar \omega\simeq
250$ MeV \cite{GR96}, and denoting the kinetic energy operator by $H_0$,
we obtain :
\begin{equation}
<s^6 [33]_{FS} | H_0 + V_{\chi} | s^6 [33]_{FS}> -  <s^4p^2 [51]_{FS} | H_0 +
V_{\chi} | s^4p^2 [51]_{FS}> \simeq 453 MeV
\end{equation}
which shows that $[51]_{FS}$ is far below the other states of Table 1. For
simplicity, here we have neglected a small difference in the confinement
potential energy between the above configurations.
 
This qualitative analysis suggests that in a more quantitative study,
where the radial dependence of the GBE interaction
is taken into account, the state $| s^4p^2
[42]_O [51]_{FS}>$ will be highly dominant and, due to
the important lowering of this state by the GBE interaction with respect
to the other states, the mixing angles with these states will be
small. That this is indeed the case, it will be proved in the section 4
below.
 
 Table 1 and the discussion above indicate that the following configurations
should be taken into account for the diagonalization of the
realistic Hamiltonian in section 4:
 
\begin{equation}
\begin{array}{ccc}
|1> &=&| s^6 [6]_O [33]_{FS} > \\
|2> &=&| s^4p^2 [42]_O [33]_{FS} > \\
|3> &=&| s^4p^2 [42]_O [51]_{FS} > \\
|4> &=&| s^4p^2 [42]_O [411]_{FS} > \\
|5> &=&| (\sqrt{\frac {5}{6}}s^52s -
\sqrt{\frac {1}{6}} s^4p^2) [6]_O [33]_{FS}>
 \\
 
\end{array}
\label{basis}
\end{equation}

A strong dominance of the  configuration $|3>$ also implies that the
one-channel approximation $\hat{A}\{NN\chi(\vec{r})\}$ is highly
inadequate for the short-range NN system. This problem will be discussed
in section 5.
 
Now we want to give a rough estimate of the interaction potential of the NN
system at zero separation distance between nucleons. We calculate
this potential in the Born-Oppenheimer (or adiabatic) approximation
defined as :
\begin{equation}
V_{NN}(R) = <H>_R - <H>_{\infty}
\label{born}
\end{equation}
where $\vec{R}$ is a collective coordinate which is the separation distance
between the two  $s^3$ nucleons, $<H>_R$ is
the lowest expectation value of the Hamiltonian describing the 6q system
at fixed $R$ and $<H>_{\infty} = 2 m_N$ for the NN problem,
i.e. the energy of two
well separated nucleons. As above we ignore the small difference
between the confinement energy of $<H>_{R=0}$ and $<H>_{\infty}$. That
this difference is small follows from the $\lambda_{i}^{c}
. \lambda_{j}^{c}$ structure of the confining interaction and from the
identity :
\begin{equation}
<[222]_c | \sum_{i<j}^{6} \lambda_{i}^{c} . \lambda_{j}^{c} | [222]_c>
= 2 <[111]_c | \sum_{i<j}^{3} \lambda_{i}^{c} . \lambda_{j}^{c} |
[111]_c> .
\end{equation}
If the space parts $[6]_O$ and $[3]_O$ contain the same
single particle state, for example an s-state, then the difference is
identically zero.
 
It has been shown by Harvey \cite{H81} that when the separation $R$ between
two $s^3$ nucleons approaches 0,  then only two types
of  6q configurations survive: $|s^6 [6]_O>$ and
$|s^4p^2 [42]_O>$. Thus in order to extract an effective NN potential
at zero separation between nucleons in adiabatic Born-Oppenheimer
approximation one has to diagonalize the Hamiltonian in the basis $|1>-|4>$.
In actual calculations in section 4  we extend the basis adding the
configuration $|5>$, which practically does not change much the result.
For the rough estimate below we take only the lowest configuration $|3>$.
One then obtains
\[
<s^4p^2 [42]_O [51]_{FS} | H_0 + V_{\chi} | s^4p^2 [42]_O [51]_{FS}> -
2 <N | H_0 + V_{\chi} | N> =
\]
\begin{equation}
\left\{
\begin{array}{ccccc}
(-100/3 + 28)C_{\chi} + 7/4 \hbar \omega & = & 280 \mbox{ MeV}, & \mbox{if} &
SI=10 \\ (-32 + 28)C_{\chi} + 7/4 \hbar \omega & = & 320 \mbox{ MeV}, &
\mbox{if} & SI=01 \\
\end{array}
\right.
\label{estim}
\end{equation}
The rough estimate (\ref{estim}) suggests that there is an effective
repulsion  of  approximately equal magnitude in the NN system in the nucleon
overlap region in both $^3S_1$ and
$^1S_0$ channels. In a more quantitative
calculation in Section 4 we find that the height of the effective core is
much larger, in particular 830 MeV for $^3S_1$  and about 1.3 GeV for $^1S_0$.
 
At this stage it is useful to compare the nature of the short-range repulsion
generated by the GBE interaction  to that produced by the OGE
interaction.
 
In the constituent quark models based on OGE the situation
is more complex.
Table 1 helps in summarizing the situation there. In this table we also
give the expectation value of the simplified chromo-magnetic interaction
\begin{equation}
V_{cm} = - C_{cm} \sum_{i<j} \lambda_{i}^{c} . \lambda_{j}^{c} \vec{\sigma}_i .
\vec{\sigma}_j
\label{vcm}
\end{equation}
in units of the  constant $C_{cm}$ (the constant $C_{cm}$ can also be
determined from the $\Delta - N$ splitting to be $C_{cm}\simeq 293/16$ MeV).
 
The expectation values of (\ref{vcm}) can be easily obtained in the CS scheme
with the help of Casimir operator formula in Appendix A and can be transformed
to FS scheme by using the unitary transformations from
the CS scheme to the FS scheme  given in Appendix B.
 
The colour-magnetic interaction pulls the configuration
$|s^4p^2[42]_O [42]_{CS}>$ down to become approximately
degenerate with $|s^6 [6]_O [222]_{CS}>$ which is pulled up.
 In a more detailed calculation
with explicit radial dependence of the colour-magnetic interaction as well
as with a Coulomb term the configuration
 $|s^6 [6]_O>$ is still the lowest one \cite{OY84,OK88}.
(With the model (\ref{vcm}) the
$\hbar \omega$ should be about 500 MeV).
Thus in the Born-Oppenheimer approximation we can roughly estimate
an effective interaction with OGE model
through the difference
 
$$
< s^6 [6]_O [222]_{CS}| H_0 + V_{cm}| s^6[6]_O [222]_{CS}> -
2 < N | H_0 + V_{cm} | N >$$
\begin{equation}
= \left\{ \begin{array}{ccc} \frac {56}{3}C_{cm} +  3/4 \hbar \omega = 717 MeV&
\mbox{if} & SI = 10 \\ 24C_{cm} +  3/4 \hbar \omega = 815 MeV & \mbox{if}
& SI = 01 \end{array}\right.
\end{equation}
 
We conclude that both the GBE and OGE models imply effective repulsion at short
range of approximately  same magnitude.
 
\section{The Hamiltonian}
 
In this section we present the GBE model \cite{GR96,GPP96} used
in the diagonalization of six-quark Hamiltonian in the basis (\ref{basis}).
The Hamiltonian reads
 
\begin{equation}
H= 6m + \sum_i \frac{\vec{p}_{i}^{2}}{2m} - \frac {(\sum_i \vec{p}_{i})^2}{12m}
+ \sum_{i<j} V_{conf}(r_{ij}) + \sum_{i<j} V_\chi(r_{ij})
\label{ham}
\end{equation}
where $m$ is the constituent quark mass; $r_{ij} = |
\vec{r}_i - \vec{r}_j |$ is the interquark distance.
 
The confining interaction is
 
\begin{equation}
 V_{conf}(r_{ij}) = -\frac{3}{8}\lambda_{i}^{c}\cdot\lambda_{j}^{c} \, C
\, r_{ij}
\label{conf}
\end{equation}
 
\noindent
where
$\lambda_{i}^{c}$  are the  SU(3)-colour matrices and C is a
parameter given below.

The spin-spin component of the GBE interaction
between the constituent quarks $i$ and $j$ reads:
 
\begin{eqnarray}
V_\chi(\vec r_{ij})
&=&
\left\{\sum_{F=1}^3 V_{\pi}(\vec r_{ij}) \lambda_i^F \lambda_j^F \right.
\nonumber \\
&+& \left. \sum_{F=4}^7 V_{\rm K}(\vec r_{ij}) \lambda_i^F \lambda_j^F
+V_{\eta}(\vec r_{ij}) \lambda_i^8 \lambda_j^8
+V_{\eta^{\prime}}(\vec r_{ij}) \lambda_i^0 \lambda_j^0\right\}
\vec\sigma_i\cdot\vec\sigma_j,
\label{VCHI}
\end{eqnarray}
 
\noindent
where $\lambda^F, F=1,...,8$ are flavour Gell-Mann matrices and
$\lambda^0 = \sqrt{2/3}~{\bf 1}$, where $\bf 1$ is the $3\times3$ unit matrix.
Thus the interaction (\ref{VCHI}) includes $\pi$,
$K$, $\eta$ and $\eta'$ exchanges. While the $\pi$,
$K$, $\eta$  mesons are (pseudo)Goldstone bosons of the spontaneously broken
$SU(3)_L \times SU(3)_R \rightarrow SU(3)_V$ chiral symmetry, the $\eta'$
(flavour singlet) is a priori not a Goldstone boson due to the axial $U(1)_A$
anomaly. In the large $N_C$ limit the axial anomaly disappears,
however, and the $\eta'$ becomes the ninth Goldstone boson of the spontaneously
broken $U(3)_L \times U(3)_R \rightarrow U(3)_V$ chiral symmetry \cite{CW80}.
Thus in the real world with $N_C = 3$ the $\eta'$ should also be taken into
account, but with parameters essentially different from $\pi$,
$K$, $\eta$ exchanges due to $1/N_C$ corrections. For the system of u and d
quarks only the $K$-exchange does not contribute.

 In the simplest case, when both the constituent quarks and mesons are
point-like particles and the boson field satisfies
the linear Klein-Gordon equation, one has the following spatial dependence
for the meson-exchange potentials \cite{GR96} :
 
\begin{equation}V_\gamma (\vec r_{ij})=
\frac{g_\gamma^2}{4\pi}\frac{1}{3}\frac{1}{4m^2}
\{\mu_\gamma^2\frac{e^{-\mu_\gamma r_{ij}}}{ r_{ij}}-4\pi\delta (\vec r_{ij})\}
,  \hspace{5mm} (\gamma = \pi, K, \eta, \eta' )
\label{POINT} \end{equation}

\noindent
where $\mu_\gamma$ are the  meson masses and $g_\gamma^2/4\pi$ are
the quark-meson coupling constants given below.
 
Eq. (\ref{POINT}) contains both the traditional long-range
Yukawa potential as well as a
$\delta$-function term. It is the latter  that is of crucial importance
for baryon spectroscopy and short-range $NN$ interaction since it has a proper
sign to provide the correct hyperfine splittings in baryons and is becoming
highly dominant at short range.
Since one deals with structured particles (both the constituent quarks and
pseudoscalar mesons) of finite extension, one must
smear out the $\delta$-function in (\ref{POINT}).
 In Ref. \cite{GPP96} a smooth Gaussian term has been employed instead of
the $\delta$-function

\begin{equation}4\pi \delta(\vec r_{ij}) \Rightarrow \frac {4}{\sqrt {\pi}}
\alpha^3 \exp(-\alpha^2(r-r_0)^2). \label{CONTACT} \end{equation}
 
\noindent
where $\alpha$ and $r_0$ are adjustable parameters.
 
The parameters of the Hamiltonian (\ref{ham}) are \cite{GPP96}:
 
$$\frac{g_{\pi q}^2}{4\pi} = \frac{g_{\eta q}^2}{4\pi} = 0.67;\,\,
\frac{g_{\eta ' q}^2}{4\pi} = 1.206$$
$$r_0 = 0.43 \, {\rm fm}, ~\alpha = 2.91 \, {\rm fm}^{-1},~~
 C= 0.474 \, {\rm fm}^{-2}, m = 340 \, {\rm MeV}. $$
\begin{equation}
 \mu_{\pi} = 139 \, {\rm MeV},~ \mu_{\eta} = 547 \, {\rm MeV},~
\mu_{\eta'} = 958 \, {\rm MeV}.
\label{PAR} \end{equation}
 
\noindent
The Hamiltonian (\ref{ham}) with the parameters (\ref{PAR}) provides a
very satisfactory description of the low-lying $N$ and $\Delta$
spectra in a fully dynamical nonrelativistic 3-body calculation \cite{GPP96}.
 
At present we are limited to use a $|s^3>$ harmonic oscillator wave function
for the nucleon in the $NN$ problem. The parametrization (\ref{PAR}) is
especially convenient for this purpose since it allows to use the $|s^3>$
as a variational ansatz. Otherwise the structure of $N$ should be more
complicated.
Indeed,  $<N|H|N>$ takes a minimal value of 969.6 MeV
at a harmonic oscillator parameter value of $\beta=0.437$ fm \cite{tetra},
i.e. only 30 MeV above the actual value in the dynamical 3-body calculation.
In this way one satisfies one of the most important constraint for the
microscopical study of the $NN$ interaction : the nucleon stability condition
\cite{OY84}
 
\begin{equation}
 \frac {\partial}{\partial b} < N | H | N > = 0 .
\label{STAB} \end{equation}
 
\noindent
The other condition, the qualitatively correct $\Delta-N$ splitting,
is also satisfied \cite{tetra}.
 
We keep in mind, however, that a nonrelativistic description of baryons cannot
be completely adequate. Within the semirelativistic description of baryons
\cite{GPPVW97} the parameters extracted from the fit to baryon masses
become considerably different and even the representation of the short-range
part of GBE (\ref{CONTACT}) has a different form. Within a semirelativistic
description the simple $s^3$ wave function for the nucleon is not adequate
anymore. All this suggests that the description of the nucleon based
 on the parameters
(\ref{PAR}) and an $s^3$ wave function is only effective. Since in
this paper we study only qualitative effects, related to the spin-flavour
structure and sign of the short-range part of GBE interaction, we consider
 the present nonrelativistic parametrization as a reasonable framework.
 
We diagonalize the Hamiltonian (\ref{ham}) in the basis (\ref{basis}). All
the necessary matrix elements are calculated with the help of the fractional
parentage technique. Some important details can be found in Appendices C and
D.

\section{Results and discussion}
In Tables II and III we present our results obtained from the
diagonalization of the
Hamiltonian (\ref{ham}) in the basis (\ref{basis}). According to the definition
of the effective potential within the Born-Oppenheimer approximation
(\ref{born}) at zero separation between nucleons
all energies presented in the Tables II and III are given
relative to two-nucleon threshold, i.e. the quantity $2<N|H|N>=1939$ MeV
has always been subtracted. In the second column we present the diagonal matrix
elements for all the states listed in the first column. In the third column we
present all the eigenvalues obtained from the diagonalization of a $5\times5$
matrix. In the fourth column  the amplitudes of all components
of the ground state are given. In agreement with Sec. 2,
one can see that the expectation value
of the configuration $|s^4p^2 [42]_O [51]_{FS}>$ given in column 2
is much lower than all
the other ones, and in particular it is about 1.5 GeV below  the expectation
value of the configuration $|s^6 [6]_O [33]_{FS}>$. The substantial
lowering of the configuration $|s^4p^2 [42]_O [51]_{FS}>$ relative to the
other ones implies that this configuration is by far the most important
component in the ground state eigenvector. The last column shows that
the probability of this configuration is 93\% both for $SI=10$ and
$SI=01$. As a consequence, the lowest eigenvalue is only about 100 MeV lower
than the expectation value of the configuration $|3>$.
 
  The main outcome is that $V_{NN}(R=0)$ is highly repulsive in both
$^3S_1$ and $^1S_0$ channels, the height being 0.830 GeV in the former
case and 1.356 GeV in the latter one.
 
  In order to see that it is the GBE interaction which is responsible for
the short
range repulsion, it is very instructive  to remove $V_\chi$ from the
Hamiltonian (\ref{ham}),  compute the ``nucleon mass" in this case,
which turns out to be $m_N=1.633$ GeV at the harmonic oscillator
parameter $\beta=0.917$ fm and diagonalize such a Hamiltonian
again in the basis (\ref{basis}). In this case the most important
configuration is $|s^6 [6]_O [33]_{FS}>$. Subtracting from the lowest
eigenvalue the ``two-nucleon energy" $2m_N=2\times1.633$ GeV one obtains
 $V^{NO ~GBE}_{NN}(R=0)=-0.197$ GeV. This soft
attraction comes from the unphysical colour Van der Waals forces related to the
pairwise confinement. The Van der Waals forces would not
appear if the basis was restricted to the $|s^6>$ state only. If
the spatially excited 3q clusters from the $s^4p^2$ configurations
were removed the Van der Waals forces  would disappear and we would arrive at
$V^{NO ~GBE}_{NN}(R=0)=0.$ Thus it is the GBE interaction which brings
about 1 GeV repulsion, consistent with the previous discussion.
 
The effective repulsion obtained above implies a strong suppression of the
$L=0$ relative motion wave function in the nucleon overlap region, as
compared to the wave function of two well separated nucleons.
 
There is another important mechanism producing additional effective repulsion
in the $NN$ system, which is related to the symmetry structure of the
lowest configuration but not related to its energy relative to the $NN$
threshold. This ``extra" repulsion, related to the ``Pauli forbidden state"
\cite{S69},  persists if any of the configurations from the $|s^4p^2>$
shell becomes highly dominant \cite{NST77}. Indeed, the $NN$ phase shift
calculated with a pure $[51]_{FS}$ state, which is projected
``by hands" (not dynamically) from
the full $NN$ state  in a toy model \cite{OY84},
shows a behaviour typical for  repulsive potentials. As a result the S-wave
NN relative motion wave function has an almost energy independent node
\cite{NOKG75}. A similar situation occurs in $^4He - ^4He$ scattering
\cite{TT65}. The only difference between this nuclear case and the NN system
is that while in the former case a configuration $s^8$ is indeed forbidden by
the Pauli principle in eight-nucleon system, the configuration $s^6$ is
allowed in a six-quark system, but is highly suppressed by dynamics, as it was
discussed above. In the OGE model this effect is absent because none of
the $[42]_O$ states is dominant \cite{OY84,OK88,HTL84}.
The existence of a strong effective repulsion, related to the energy balance
in the adiabatic approximation, as in our case, suggests, however, that the
amplitude of the oscillating NN wave function at short distance will be
strongly suppressed.
 
To illustrate this discussion we project the lowest eigenvector in Table II
 onto the $NN$ and $\Delta \Delta$ channels. The projection onto any
baryon-baryon channel $B_1B_2$ is defined as follows \cite{KNO91,GNO93}
 
\begin{equation}
 \Psi_{B_1B_2}(\vec r) = \sqrt{\frac {6!}{3! 3! 2}}
< B_1(1,2,3) B_2(4,5,6) | \Psi (1,2,...,6)>,
\label{PROJ} \end{equation}
 
\noindent
where $\Psi (1,2,...,6)$ is a fully antisymmetric 6q wave function,
which in the present case is represented by the eigenvector in Table II,
and $B_1(1,2,3)$ and $B_2(4,5,6)$ are intrinsic  baryon wave functions.
 
In order to calculate (\ref{PROJ}) we need a $``3+3"$ expansion of each
state in the basis (\ref{basis}). The corresponding $``3+3"$ decomposition
of each state can be found in \cite{GNO93} in the $CS$ coupling scheme.
To use it here one needs the unitary matrix
 from the $CS$ basis to the $FS$ one. This matrix can be found in
Appendix B.
 
In Fig. 1 we show the projections (\ref{PROJ}) onto the $NN$ and $\Delta
\Delta$ channels in the $^3S_1$ partial $NN$ wave at short range. In fact, such
projections can be shown for other channels too as e.g. $NN^*$, $N^*N^*$,...
some of them being  not small. Note that our six-quark wave function,
calculated at short range only, was normalized to 1. Hence, we cannot show the
suppression of the $NN$ projection in the nucleon overlap region as compared
to the wave function of the well separated nucleons, discussed above.
This can only be seen from $\Psi_{NN}(\vec r)$ obtained in
 dynamical calculations
including not only the short-range 6q configurations, like in the present
paper, but also the basis states representing the middle- and long distances
in the $NN$ system.
 
In Fig. 1 one observes  a nodal behaviour of both $\Psi_{NN}(\vec r)$
and $\Psi_{\Delta \Delta}(\vec r)$ at short range. Also
$\Psi_{\Delta \Delta}(\vec r)$ is essentially larger at short range than
$\Psi_{NN}(\vec r)$. At large distances only $\Psi_{NN}(\vec r)$ will survive.
This nodal behaviour is related to the fact that the configuration $|3>$
is highly dominant. In the case of any configuration $s^4p^2$ or $s^52s$ from
the N=2 shell, the relative motion of two $s^3$ clusters (e.g. $NN$ and $\Delta
\Delta$) is described by a nodal wave function.
 
Now we want to discuss the question which type of $NN$ potential would
be equivalent to the short-range picture obtained above. If one considers
the effect of the short-range dynamics on the $NN$ phase
shifts, in a limited energy interval the phase shifts in both
 $^3S_1$ and $^1S_0$ partial waves can be simulated by
strong repulsive core potentials or by ``deep attractive potentials with
forbidden states" \cite{NOKG75}. The latter potentials are in fact
supersymmetric partners of the former ones \cite{MR88}.
 
If, on the other hand, one considers the effect of the short-range dynamics
on the structure of the wave function at short range, it is difficult
to construct a potential which would be adequate. For example, a repulsive
core potential produces a wave function which is indeed suppressed at
short range,
but does not have any nodal structure. If one takes, instead, a ``deep
attractive potential with forbidden state", one obtains a nodal behaviour,
but the wave function is not suppressed at short range (i.e. the amplitude
left to the node is a very large one). As a direct consequence,
the latter potential produces a very rich high-momentum component,
which is in contradiction with the deuteron electromagnetic form factors
\cite{GBK89}. The high-momentum component, implied by a ``very soft node",
like in our case, will be much smaller and closer to that one obtained
from the potentials with strong repulsive core.
 
We also see large projections onto other $B_1B_2$ channels (exemplified by
the $\Delta \Delta$ channel in Fig. 1). These components cannot be taken
into account in any simple $NN$ potential, in principle. Thus, if we
are interested in effects, related to the short-range $NN$ system, there is
no way, other than to consider the full 6q wave function in this region.

\section{Why the single channel resonating group method ansatz is not
adequate ?}
In this section we show that the currently used one-channel resonating
group method (RGM) ansatz for the two-nucleon wave function is not adequate
in a study of the short-range $NN$ interaction with the chiral quark model.

In the
one-channel RGM approximation the 6q wave function has the form
\begin{eqnarray}
\psi & = & \hat{A}\{N(1,2,3)N(4,5,6)\chi(\vec{r})\}, \label{WF}\\
\nonumber
\hat{A}&=& \frac{1}{\sqrt{10}}(1-9 \hat{P}_{36}), \\
\nonumber
\vec{r}&=& \frac{\vec{r}_1+\vec{r}_2+\vec{r}_3}{3} -
\frac{\vec{r}_4+\vec{r}_5+\vec{r}_6}{3}.
\end{eqnarray}
This is reasonable in the case where the short-range quark
dynamics is described in
terms of the OGE interaction. In this case the addition of new channels,
orthogonal to (\ref{WF}), does not change considerably the full wave function
in the nucleon overlap region. This
is not the case for the chiral constituent quark model, where the short-range
quark dynamics is due to the GBE interaction. To have a better
insight why (\ref{WF}) is a poor approximation in the present case, we begin
with the explanation why (\ref{WF}) is reasonable for OGE model \cite{OK88} .
 
To this end, it is very convenient to use the six-quark shell model
basis for the NN
function in the nucleon overlap region \cite{KNO91,OK88,OOAY82}. Such a basis
is much more flexible than (\ref{WF}). Diagonalizing a Hamiltonian comprising
OGE and a confining interaction in the harmonic oscillator basis up to two
excitation quanta, one can obtain the 6q wave function in the form \cite{OK88}
\begin{equation}
\psi = C_0 |s^6> + \sum_{\alpha} C_{\alpha} |\alpha>,
\label{sum}
\end{equation}
where $\alpha$ lists all possible configurations in the N=2 shell : $
[6]_O [222]_{CS}$, $[42]_O [42]_{CS}$, $[42]_O [321]_{CS}$, $[42]_O
[222]_{CS}$, $[42]_O [3111]_{CS}$, $[42]_O [21111]_{CS}$. With the OGE
interaction, the CS coupling scheme based on the chain $SU(6)_{CS} \supset
SU(3)_{C} \times SU(2)_S$ is more convenient.
It has been found that there are a few most important
configurations - $ |s^6 [6]_O[222]_{CS}>, |s^4p^2 [42]_O [42]_{CS}>, |s^4p^2
[42]_O [321]_{CS}>, |(\sqrt{5/6} s^52s - \sqrt{1/6} s^4p^2) [6]_O [222]_{CS}>$
- with sizeable amplitudes $C_{\alpha}$ \cite{OK88,GNO93,OOAY82}.
 
Now, let us expand the RGM wave function (\ref{WF}) in the shell model basis.
For that purpose, the trial function $\chi(\vec{r})$ in (\ref{WF}) should be
expanded in a harmonic oscillator basis too
\begin{equation}
\chi_{L=0}(\vec{r}) = \sum_{N=0,2,4,...} <\chi_{L=0} | \phi_{NS}>
\phi_{NS}(\vec{r}),
\label{exp}
\end{equation}
where $\phi_{NS}(\vec{r})$ is a harmonic oscillator state with N quanta and
$L=0$. Thus in the ansatz (\ref{WF}) the variational coefficients based on
the expansion (\ref{exp}) are $<\chi_{L=0} | \phi_{NS}>$. The last step is to
use the expressions (21) and (22) of Ref. \cite{GNO93} for
$\hat{A}\{N(1,2,3)N(4,5,6)\phi_{0S}(\vec{r})\}$,
$\hat{A}\{N(1,2,3)N(4,5,6)\phi_{2S}(\vec{r})\}$, written in the shell
model basis. These are transformations from one basis to another and do not
depend on the 6q dynamics. If it turns out that for a given Hamiltonian the
variational coefficients $C_{\alpha}$ in (\ref{sum}) are close to the
algebraical ones
\mbox{$<\alpha | \hat{A}\{N(1,2,3)N(4,5,6)\phi_{2S}(\vec{r})\}>$}, then one can
conclude that (\ref{WF}) is a good approximation for the
variational solution (\ref{sum}) . If not, the variational ansatz (\ref{WF})
is poor and
other channels, not equivalent to (\ref{WF}), should be added (e.g.
$\hat{A}\{N^*N\chi^*(\vec{r})\}, \hat{A}\{N^*N^*\chi^{**}(\vec{r})\}$,...).
For the OGE model it is found that indeed the variational coefficients
$C_{\alpha}$ in (\ref{sum}) are very close to the algebraical ones
\cite{OK88} (see also \cite{GNO93}).
 
Let us now turn to the analysis of the results of Sec. 4 based on the GBE
interaction. Using the unitary transformation from the CS to FS scheme, given
in Appendix B, one can rewrite
Eqs.(21) and (22) of Ref.\cite{GNO93} as :
\begin{eqnarray}
\hat{A}\{N(1,2,3)N(4,5,6)\phi_{0s}(\vec{r})\}_{SI=10} =
\sqrt{\frac {10}{9}} |s^6[6]_O [33]_{FS}>, \\
\hat{A}\{N(1,2,3)N(4,5,6)\phi_{2s}(\vec{r})\}_{SI=10} = \frac {3\sqrt{2}}{9}
|(\sqrt{\frac {5}{6}} s^52s - \sqrt{\frac {1}{6}} s^4p^2) [6]_O [33]_{FS}>
  \nonumber\\
-\frac {4 \sqrt{2}}{9} | s^4p^2 [42]_O [33]_{FS}> -
\frac {4 \sqrt{2}}{9} | s^4p^2 [42]_O
[51]_{FS}>. \label{unit}
\end{eqnarray}
 
\noindent
From the expression (\ref{unit}) we see that the relative amplitudes of the
states $|5>$, $|2>$ and $|3>$ are in the ratio
\begin{equation}
|5> : |2> : |3> = 3 : -4 : -4
\end{equation}
and the amplitude of the state $|4>$ is zero. The
diagonalization of the Hamiltonian made in the previous section gives
\begin{equation}
|5> : |2> : |3> : |4> \simeq 0.06 : 0.08 : -0.96 : 0.20 ,
\end{equation}
Therefore the ansatz (\ref{WF}) is completely inadequate in the nucleon
overlap region and the incorporation of additional
channels is required in RGM calculations.
 
\section{Summary}
 
In the present paper we have calculated an adiabatic $NN$ potential
at zero separation between nucleons in the framework of a chiral
constituent quark model, where the constituent quarks interact via
pseudoscalar meson exchange. Diagonalizing a Hamiltonian in a basis
consisting of the most important 6q configurations in the nucleon
overlap region, we have found a very strong effective repulsion of the order
 of 1 GeV in both $^3S_1$ and $^1S_0$ $NN$ partial waves. Due to the specific
flavour-spin symmetry of the Goldstone boson exchange interaction the
configuration $|s^4p^2 [42]_O [51]_{FS}>$ becomes highly dominant
at short range. As a consequence, the projection of the full 6q wave function
onto the $NN$ channel should have a node at short range in both
$^3S_1$ and $^1S_0$ partial waves. The amplitude of the oscillation left
to the node should be strongly suppressed as compared to the wave function
of two well separated nucleons.
 
We  have also found that due to the strong dominance of the configuration
$|s^4p^2 [42]_O [51]_{FS}>$ the commonly used one-channel RGM ansatz is
a very poor approximation to the 6q wave function in the nucleon overlap
region.
 
Thus, within the chiral constituent quark model one has all the necessary
ingredients to understand microscopically the $NN$ interaction. There
appears strong effective short-range repulsion from the same part of Goldstone
boson exchange which also produces hyperfine splittings in baryon
spectroscopy. The long- and middle-range attraction in the $NN$ system
is automatically implied by the Yukawa part of pion exchange and
two-pion (or $\sigma$) exchanges between quarks belonging to different
nucleons. With this first encouraging result, it might be
worthwhile to perform a more elaborate calculation of $NN$ system and
other baryon-baryon systems within the present framework.
 
\appendix
\section{}
The expectation value of the operators (\ref{opFS}) and (\ref{vcm}),
displayed in Table \ref{expectation}, are calculated with the following
formulae:
\begin{equation}
<\sum_{i<j} \lambda_i . \lambda_j \vec{\sigma}_i . \vec{\sigma}_j> = 4
C_{2}^{SU(6)} - 2 C_{2}^{SU(3)} -
\frac{4}{3} C_{2}^{SU(2)} - 8N
\end{equation}
where N is the number of particles, here N=6, and $C_{2}^{SU(n)}$ is the
Casimir operator eigenvalues of $SU(n)$ which can be derived from the
expression :
\begin{eqnarray}
C_{2}^{SU(n)} = \frac{1}{2} [f_1'(f_1'+n-1) + f_2'(f_2'+n-3) + f_3'(f_3'+n-5)
 \nonumber \\
+f_4'(f_4'+n-7) + ... + f_{n-1}'(f_{n-1}'-n+3) ] - \frac{1}{2n}
(\sum_{i=1}^{n-1} f_i')^2
\label{casimir}
\end{eqnarray}
where $f_i'= f_i-f_n$, for an irreductible representation given by the
partition $[f_1,f_2,...,f_n]$.
 
\section{}
This appendix reproduces transformations, derived elsewhere, from the CS
coupling scheme to the FS coupling scheme, or vice versa, related to the
orbital symmetries $[6]_O$ and $[42]_O$, appearing in the basis vectors
(\ref{basis}).
 
For the $[6]_O$ symmetry one obviously has :
\begin{equation}
[6]_O [33]_{FS} = [6]_O [222]_{CS}
\end{equation}
either for IS=01 or 10.
 
For the $[42]_O$ symmetry, sector IS=01, the Table \ref{CS} reproduces Table 7
of Ref.
\cite{St89} with a phase change in columns 3 and 5, required by consistency
with Ref. \cite{GNO93}.
 
In this Table, the column headings are
\begin{equation}
\begin{array}{ccc}
\psi_{1}^{CS} $ = $ [42]_O [42]_{CS} \\
\psi_{2}^{CS} $ = $ [42]_O [321]_{CS} \\
\psi_{3}^{CS} $ = $ [42]_O [3111]_{CS} \\
\psi_{4}^{CS} $ = $ [42]_O [222]_{CS} \\
\psi_{5}^{CS} $ = $ [42]_O [21111]_{CS}
\end{array}
\end{equation}
For the $[42]_O$ symmetry, sector IS=10, we reproduce in Table \ref{Ping}
the corresponding Table from Ref.\cite{PWG97} by interchanging rows with
columns and then reorder the rows. In this case, the notation is
\begin{equation}
\begin{array}{ccc}
\bar{\psi}_{1}^{CS} & = & [42]_O [411]_{CS} \\
\bar{\psi}_{2}^{CS} & = & [42]_O [33]_{CS} \\
\bar{\psi}_{3}^{CS} & = & [42]_O [2211]_{CS}^{1} \\
\bar{\psi}_{4}^{CS} & = & [42]_O [2211]_{CS}^{2} \\
\bar{\psi}_{5}^{CS} & = & [42]_O [1^6]_{CS}
\end{array}
\end{equation}
The upper indices 1 and 2 take into account the two representations $[2211]$
appearing in the inner product $[222] \times [33]$.
 
\section{}
The calculation of the matrix elements of the Hamiltonian (\ref{ham}) is
based on the fractional parentage (cfp) technique described in Ref.
\cite{H81}. For details, see also Ref. \cite{book}, chapter 10. In dealing
with n particles the matrix elements of a symmetric two-body operator between
totally (symmetric or) antisymmetric states $\psi_n$ and $\psi_n'$ reads
\begin{equation}
<\psi_n | \sum_{i<j} V_{ij} | \psi_n' > = \frac{n(n-1)}{2} <\psi_n |
 V_{n-1,n} | \psi_n' >
\end{equation}
The matrix elements of $V_{n-1,n}$ are calculated by expanding $\psi_n$ and
$\psi_n'$ in terms of products of antisymmetric states of the first n-2
particles $\psi_{n-2}$ and of the last pair $\phi_2$
\begin{equation}
\psi_n = \sum_{\alpha \beta} P_{\alpha\beta} \psi_{n-2}(\alpha) \phi_2(\beta)
\end{equation}
with $\alpha, \beta$ denoting the possible structures of $\psi_{n-2}$ and
$\phi_2$ and $P_{\alpha\beta}$ the products of cfp coefficients in the
orbital, spin-flavour and colour space states. In practical calculations, the
colour space cfp coefficients are not required. The orbital cfp are taken
from Ref. \cite{SW88}, Tables 1 and 2 by using the replacement $r^4l^2
\rightarrow s^4p^2$ and $r^5l \rightarrow s^5p$. The trivial ones are equal
to one. The flavour-spin cfp for IS=01 are identical to the
$\bar{K}$-matrices of Table 1 Ref. \cite{St89} with $[42]_S [33]_F$ in the
column headings. For IS=10 they are the same as for IS=01 but the column
headings is $[42]_F [33]_S$ instead of $[42]_S [33]_F$ as above, and this is
due to the commutativity of inner products of $S_n$ (see for example Ref.
\cite{book}). The cfp used in the OC coupling are from Ref. \cite{SW88} Table
3, for $[42]_O \times [222]_C \rightarrow [3111]_{OC}$ and Table 5 of Ref.
\cite{St89} for $[42]_O \times [222]_C \rightarrow [222]_{OC}$ and $[42]_O
\times [222]_C \rightarrow [21111]_{OC}$.
 
In this way, after decoupling all degrees of freedom one can integrate out in
the colour, spin and flavour space. The net outcome of this algebra is that
any six-body matrix element becomes a linear combination of two-body orbital
matrix elements, $<V_{\pi}>, <V_{\eta}>$ and $<V_{\eta'}>$. The coefficients
of $<V_{\pi}>$ are the same for IS=01 and 10, but the coefficients of
$<V_{\eta}>$ are usually different. In both cases the coefficients of
$<V_{\eta'}>$ are two times those of $<V_{\eta}>$. We found that the two-body
GBE matrix elements satisfy the relations
$<V_{\pi}>
\simeq <V_{\eta}>$ and
$<V_{\eta'}>
\simeq 2<V_{\pi}>$. As an example, in Table \ref{OME} we show the matrix
elements obtained for IS=01. Except for $<ss | V |pp>$ and $<s2s | V | pp>$
they are all negative, i.e. carry the sign of Eq. (\ref{opFS}).
 
In a harmonic oscillator basis the confinement potential matrix elements can
be performed analytically. As an illustration, in Appendix D, we reproduce
the results for all configurations required in these calculations.
 
Finally, the kinetic energy matrix elements can be calculated as above, by
writting the relative kinetic energy operator as a two-body operator
\begin{equation}
T = \sum_i \frac{p_{i}^{2}}{2m} - \frac{1}{12} (\sum_i \vec{p}_i)^2 =
 \sum_{i<j} T_{ij}
\end{equation}
with
\begin{equation}
T_{ij} = \frac{1}{12m} (p_{i}^{2} + p_{j}^{2}) - \frac{1}{6m} \vec{p_i} .
\vec{p_j}
\end{equation}
Alternatively we can use an universal formula for the kinetic energy of
harmonic oscillator states
\begin{equation}
K.E. = \frac{1}{2} [N + \frac{3}{2} (n-1) ] \hbar \omega + \frac{3}{4} \hbar
\omega
\end{equation}
where N is the number of quanta and n the number of particles. The last
term is the kinetic energy of the center of mass.
 
\section{}
We work with the following single particle harmonic oscillator states :
\begin{eqnarray}
|s> & = & \pi^{-3/4} \beta^{-3/2} \exp{(-r^2/2\beta^2)} \\
|p>_m & = & 8^{1/2} 3^{-1/2} \pi^{-1/4} \beta^{-5/2} r \exp{(-r^2/2\beta^2)}
\, Y_{1m} \\
|2s> & = & 2^{1/2} 3^{-1/2} \pi^{-3/4} \beta^{-3/2}
(\frac{3}{2} -\frac{r^2}{\beta^2} ) \exp{(-r^2/2\beta^2)}
\end{eqnarray}
In this basis the two-body matrix elements of the confining potential
$V^c=Cr$ of Eq.(\ref{conf}) are
\begin{eqnarray}
<ss | V^c | ss> & = & \sqrt{\frac{2}{\pi}} 2 C \beta \\
<sp | V^c | sp> & = & \sqrt{\frac{2}{\pi}} \frac{7 C \beta}{3} \\
<sp | V^c | ps> & = & -\sqrt{\frac{2}{\pi}}  \frac{ C \beta}{3} \\
<ss | V^c | (pp)_{L=0}> & = & -\sqrt{3} <sp | V^c | ps> \\
<(pp)_{L=0} | V^c | (pp)_{L=0}> & = & \sqrt{\frac{2}{\pi}}  
\frac{5 C \beta}{2} \\
<s2s | V^c | s2s> & = & \sqrt{\frac{2}{\pi}} \frac{31 C \beta}{12} \\
<ss | V^c | s2s> & = & -\sqrt{\frac{1}{3\pi}}  C \beta \\
<s2s | V^c | (pp)_{L=0}> & = & -\frac{1}{\sqrt{\pi}} \frac{C \beta}{2}
\end{eqnarray}

\begin{table}
\renewcommand{\arraystretch}{1.5}
\caption[Expectation values]{\label{expectation} Expectation values of the
operators defined by Eqs. (\ref{opFS}) and (\ref{vcm}) for all compatible
symmetries
$[f]_O [f]_{FS}$ in the IS=(01) and (10) sector. $<V_{\chi}>$ is in units of
$C_{\chi}$ and
$<V_{cm}>$ in units of $C_{cm}$.}
\begin{tabular}{|c|c|c|c|c}
      & \multicolumn{2}{c|}{I=0 S=1} &
\multicolumn{2}{c}{I=1 S=0} \\
$[f]_o [f]_{FS}$  & $<V_{\chi}>$ & $<V_{cm}>$ & $<V_{\chi}>$
& $<V_{cm}>$ \\
\tableline
$[6]_o [33]_{FS}$  & -28/3 & 8/3 & -8 & 8 \\
$[42]_o [33]_{FS}$  & -28/3 & -26/9 & -8 & -4/3 \\
$[42]_o [51]_{FS}$  & -100/3 & 16/9 & -32 & 16/9 \\
$[42]_o [411]_{FS}$  & -28/3 & 20/9 & -8 & 44/9 \\
$[42]_o [321]_{FS}$  & 8/3 & -164/45 & 4 & 232/45 \\
$[42]_o [2211]_{FS}$  & 68/3 & -62/15 & 60 & 42/5 \\
\end{tabular}
\end{table}
 
\begin{table}
\renewcommand{\arraystretch}{1.5}
\parbox{18cm}{\caption[Results IS=01]{\label{value01} Results of the
diagonalization of  the Hamiltonian
(\ref{ham}) for IS=(01). Column 1 - the basis states,
column 2 - diagonal matrix elements (GeV), column 3 - eigenvalues (GeV)
for a $5 \times 5$ matrix, column 4 - components of
the lowest state. The results are for $\beta = 0.437$ fm. In columns 2 and 3,
the quantity $2 m_N = 1.939$ GeV is subtracted.}}
\\[4mm]
\begin{tabular}[t]{|c|c|c|c|}
 state & $1 \times 1$ & $5 \times 5$ & lowest state \\
       &  &  & components \\ \hline
$| s^6 [6]_o [33]_{FS} >$ & 2.346 & 0.830 & -0.14031 \\
$| s^4p^2 [42]_o [33]_{FS} >$ & 2.824 & 1.323 & 0.07747 \\
$| s^4p^2 [42]_o [51]_{FS} >$ & 0.942 & 2.693 & -0.96476 \\
$| s^4p^2 [42]_o [411]_{FS} >$ & 2.949 & 3.049 & 0.20063 \\
$| (\sqrt{5/6} s^{5}2s - \sqrt{1/6}s^4p^2  [6]_o [33]_{FS} >$ & 3.011 &
4.169 & 0.05747 \\
\end{tabular}
\end{table}
 
\begin{table}
\renewcommand{\arraystretch}{1.5}
\caption{\label{value10} Same as Table 2 but
for IS=(10).}
\begin{tabular}[t]{|c|c|c|c|}
 state & $1 \times 1$ & $5 \times 5$ & lowest state \\
       &  &  & components \\ \hline
$| s^6 [6]_o [33]_{FS} >$ & 2.990 & 1.356 & -0.12195 \\
$| s^4p^2 [42]_o [33]_{FS} >$ & 3.326 & 1.895 & 0.08825 \\
$| s^4p^2 [42]_o [51]_{FS} >$ & 1.486 & 3.178 & -0.96345 \\
$| s^4p^2 [42]_o [411]_{FS} >$ & 3.543 & 3.652 & -0.21644 \\
$| (\sqrt{5/6} s^{5}2s - \sqrt{1/6}s^4p^2  [6]_o [33]_{FS} >$ & 3.513 &
4.777 & 0.04756 \\
\end{tabular}
\end{table}

\begin{table}
\renewcommand{\arraystretch}{1.5}
\caption[CSTS]{\label{CS} The unitary transformation between the CS and FS
basis vectors of orbital symmetry $[42]_o$, isospin I=0, and spin S=1.}
\begin{tabular}{c|ccccc}
 & $\psi_{1}^{CS}$ & $\psi_{2}^{CS}$ & $\psi_{3}^{CS}$ & $\psi_{4}^{CS}$ &
$\psi_{5}^{CS}$ \\ \hline
$[42]_o [33]_{FS}$ & $\frac{9\sqrt{5}}{36}$ & $-\frac{8\sqrt{5}}{36}$ &
$\frac{5\sqrt{2}}{36}$ & $\frac{11}{36}$ & $\frac{20}{36}$ \\
$[42]_o [51]_{FS}$ & $\frac{9\sqrt{5}}{45}$ & $-\frac{8\sqrt{5}}{45}$ &
$\frac{5\sqrt{2}}{45}$ & $-\frac{25}{45}$ & $-\frac{25}{45}$ \\
$[42]_o [411]_{FS}$ & $\frac{9\sqrt{10}}{180}$ & $-\frac{8\sqrt{10}}{180}$ &
$-\frac{170}{180}$ & $-\frac{25\sqrt{2}}{180}$ & $\frac{20\sqrt{2}}{180}$ \\
$[42]_o [2211]_{FS}$ & $\frac{11}{20}$ & $\frac{8}{20}$ &
$-\frac{\sqrt{10}}{20}$ & $\frac{5\sqrt{5}}{20}$ & $-\frac{4\sqrt{5}}{20}$ \\
$[42]_o [321]_{FS}$ & $-\frac{18}{45}$ & $-\frac{29}{45}$ &
$-\frac{2\sqrt{10}}{45}$ & $\frac{10\sqrt{5}}{45}$ & $-\frac{8\sqrt{5}}{45}$
\end{tabular}
\end{table}

\begin{table}
\renewcommand{\arraystretch}{1.5}
\caption[Ping]{\label{Ping} Same as Table \ref{CS} but for S=0, I=1.}
\begin{tabular}{c|ccccc}
 & $\bar{\psi}_{1}^{CS}$ & $\bar{\psi}_{2}^{CS}$ & $\bar{\psi}_{3}^{CS}$ &
$\bar{\psi}_{4}^{CS}$ &
$\bar{\psi}_{5}^{CS}$ \\ \hline
$[42]_o [33]_{FS}$ & $\sqrt{\frac{25}{72}}$ & $-\sqrt{\frac{25}{144}}$ &
$-\sqrt{\frac{49}{144}}$ & $-\sqrt{\frac{1}{36}}$ & $-\sqrt{\frac{1}{9}}$ \\
$[42]_o [51]_{FS}$ & $\sqrt{\frac{2}{9}}$ & $-\sqrt{\frac{1}{9}}$ &
$\sqrt{\frac{1}{9}}$ & $\sqrt{\frac{4}{9}}$ & $\sqrt{\frac{1}{9}}$ \\
$[42]_o [411]_{FS}$ & $\sqrt{\frac{1}{36}}$ & $\sqrt{\frac{25}{72}}$ &
$-\sqrt{\frac{25}{72}}$ & $\sqrt{\frac{1}{18}}$ & $\sqrt{\frac{2}{9}}$ \\
$[42]_o [2211]_{FS}$ & $-\sqrt{\frac{9}{40}}$ & $-\sqrt{\frac{1}{80}}$ &
$-\sqrt{\frac{9}{80}}$ & $\sqrt{\frac{9}{20}}$ & $-\sqrt{\frac{1}{5}}$ \\
$[42]_o [321]_{FS}$ & $-\sqrt{\frac{8}{45}}$ & $-\sqrt{\frac{16}{45}}$ &
$-\sqrt{\frac{4}{45}}$ & $-\sqrt{\frac{1}{45}}$ & $\sqrt{\frac{16}{45}}$ \\
\end{tabular}
\end{table}
 
\begin{table}
\renewcommand{\arraystretch}{1.5}
\parbox{18cm}{\caption[One-meson exchange matrix elements]{\label{OME} All
one-meson exchange two-body matrix elements (GeV) for the sector IS=01
evaluated at $\beta = 0.437$ fm. The remaining one is
$<sp | V_F | ps> = - <ss | V_F |(pp)_{L=0}>/\sqrt{3}$.}}
\\[4mm]
\begin{tabular}[t]{|c|l|l|l|}
Two-body matrix  & $F=\pi$ & $F=\eta$ & $F=\eta'$ \\
 elements & & & \\ \hline
$<ss | V_F | ss>$ & -0.108357 & -0.104520 & -0.189153 \\
$<ss | V_F | (pp)_{L=0}>$ & 0.043762 & 0.042597 & 0.076173 \\
$<sp | V_F | sp>$ & -0.083091 & -0.079926 & -0.145175 \\
$<(pp)_{L=0} | V_F | (pp)_{L=0}>$ & -0.081160 & -0.078594 & -0.142205 \\
$<s2s | V_F | s2s>$ & -0.069492 & -0.066963 & -0.121701 \\
$<ss | V_F | s2s>$ & -0.030945 & -0.030121 & -0.053863 \\
$<s2s | V_F | (pp)_{L=0}>$ & 0.033309 & 0.031753 & 0.057499 \\
\end{tabular}
\end{table}

\bigskip
\bigskip
\bigskip
{\center{\bf Figure Captions}}

\bigskip
{\center{\bf Fig. 1}}

Projections of the lowest eigenvector in Table II onto $NN$ and $\Delta
\Delta$ channels (in arbitrary units). 
 
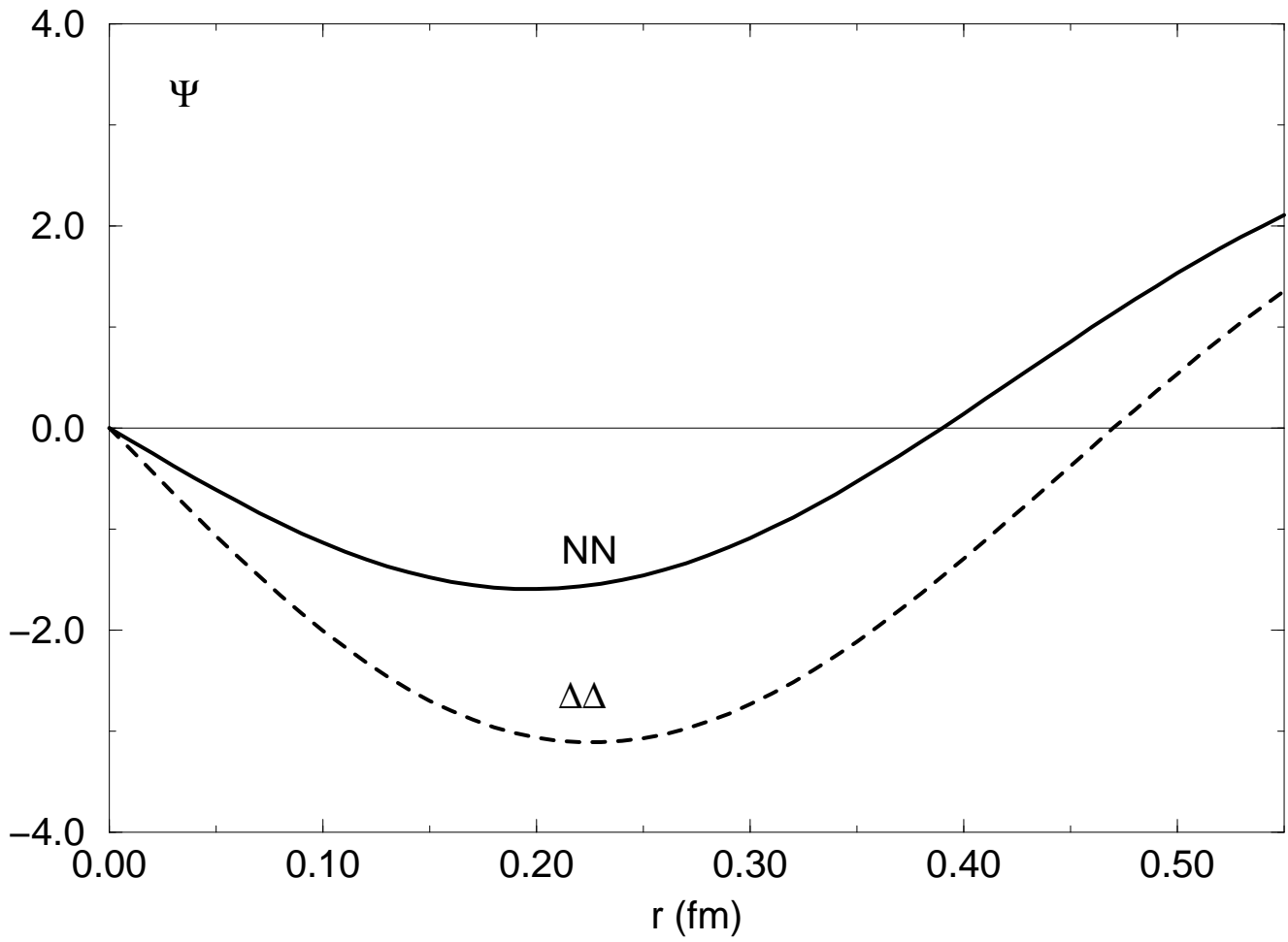
\end{document}